\documentclass[12pt,twoside,draft]{article}
\usepackage{amsmath}
\usepackage{amsfonts,amssymb,latexsym}
\usepackage{amscd}
\usepackage{eucal,mathrsfs}
\usepackage{amsthm}
\newtheorem{theorem}{\bf{Theorem}}[section]

\newtheorem{proposition}[theorem]{\bf{Proposition}} 
 
\newtheorem{definition}[theorem]{\bf{Definition}}

\theoremstyle{remark}

\newtheorem{remark}{Remark}

\usepackage{graphics}
\numberwithin{equation}{section}
\def\Bbb{\mathbb}
\newcounter{contador}

\def\logo{\raisebox{-10.5\p@}{\hb@xt@85\p@{\includegraphics{gft.eps}\hfil}}}

\def\un{1\kern-3pt \rm I}
\def\ptoday{{\ifcase\month 
\or January, \or February, \or March, \or April,\or May, 
\or June, \or July, \or August, \or September, \or October, 
\or November, \or December,\fi\ \number \year}}

\newcommand{\oR}{{\mathbb R}}

\newcommand{\oC}{{\mathbb C}}

\input epsf
\usepackage[dvips]{graphicx}

\def\dj{\hbox{d\kern-0.347em \vrule width 0.3em height 1.252ex depth
-1.21ex \kern 0.051em}}

\begin{document}

\title{\rm A New Derivation of the CPT and Spin-Statistics \\[3mm]
Theorems in Non-Commutative Field Theories}

\author{\normalsize{\bf Daniel H.T. Franco}\footnote{On leave from the
Centro de Estudos de F\'\i sica Te\'orica, Belo Horizonte, MG,
Brasil.} \footnote{e-mail: dhtf@terra.com.br}\,\,$\textsuperscript{$(a)$}$ and 
{\bf Caio M.M. Polito}$\textsuperscript{$(b)$}$ \\
\\
{\normalsize {$(a)$} {\em Universidade Federal do Esp\'\i rito Santo (UFES)}}\\
{\normalsize {\em Departamento de F\'\i sica -- Campus Universit\'ario de Goiabeiras}}\\
{\normalsize {\em CEP:29060-900 -- Vit\'oria -- ES -- Brasil}}
\\
\\
{\normalsize {$(b)$} {\em Centro Brasileiro de Pesquisas F\'{\i}sicas -- (CBPF)}} \\ 
{\normalsize {\em Coordena\c c\~ao de Campos e Part\'{\i}culas -- (CCP)}} \\
{\normalsize {\em Rua Dr. Xavier Sigaud 150 -- Urca}}\\
{\normalsize {\em CEP:22290-180 -- Rio de Janeiro -- RJ -- Brasil}}}

\date{\ptoday}

\maketitle

\vspace{-1cm}


\begin{abstract}
We propose an alternative axiomatic description for non-commutative field theories
(NCFT) based on some ideas by Soloviev to nonlocal quantum fields. The local
commutativity axiom is replaced by the weaker condition that the fields commute at
sufficiently large spatial separations, called asymptotic commutati\-vi\-ty, formulated
in terms of the theory of analytic functionals.
The question of a possible violation of the CPT and Spin-Statistics theorems caused by
nonlocality of the commutation relations $[\widehat{x}_\mu,\widehat{x}_\nu]=i\theta_{\mu\nu}$
is investigated. In spite of this inherent nonlocality, we show that the modification
aforementioned is sufficient to ensure the validity of these theorems for NCFT. We
restrict ourselves to the simplest model of a scalar field in the case of only space-space
non-commutativity.
\end{abstract}

\,\,\,{Keywords: Non-commutative theory, analytic functionals, axiomatic field theory.}

\section{Introduction}
\hspace*{\parindent}
In recent years a considerable effort has been made to clarify the structural
aspects of NCFT. The first paper on quantum field theory by exploring the
non-commutativity of a space-time manifold was published by Snyder, in 1947~\cite{Sny},
who used this idea to give a solution for the problem of ultraviolet divergences which
had plagued quantum field theories from very beginning. Since then, due to the success
of the renormalization theory, this subject was abandoned. Only recently the plan of
investigating field theories on non-commutative space-times has been revived. In a
fundamental paper Doplicher-Fredenhagen-Roberts~\cite{DFR} have shown that a model quantum
space-time can be described by a non-commutative algebra whose commutation relations do
imply uncertainty relations motivated by Heisenberg's uncertainty principle and
by Einstein's theory. Later, in a different context, NCFT appear directly related with
the string theory~\cite{SeiWit}: the non-commutative Yang-Mills theory can be seen as a
vestige, in the low-energy limit, of open strings in the presence of a constant magnetic
field, $B_{\mu\nu}$ (for a review see~\cite{DouNe,Sza}).

From an axiomatic standpoint, a language has been developed which, in principle,
ought to enable one to extend the Wightman axioms to this context. \'Alvarez-Gaum\'e
and V\'asquez-Mozo~\cite{AGVM} have recently taken the first step to examine general
properties of NCFT within the axiomatic framework by modifying the standard
Wightman axioms. By using as guiding principles the breaking of Lorentz symmetry down to
the subgroup $O(1,1) \times SO(2)$, which leaves invariant the commutation relations for
the coordinate operators $[\widehat{x}_\mu,\widehat{x}_\nu]=i \theta_{\mu\nu}$, and the
relaxation of local commutativity to make it compatible with the causal structure of the
theory, described by the light-wedge associated with the $O(1,1)$ factor of the kinematical
symmetry group, they have demonstrated the validity of the CPT theorem for NCFT. As it was
stressed in~\cite{AGVM}, a source of difficulties in formulating NCFT which satisfy the
adapted axioms has been the very harmful UV/IR mixing, which is probably the most surprising
feature of these theories. The existence of hard infrared singularities in the non-planar
sector of the theory, induced by uncancelled quadratic ultraviolet divergences, can result
in two kinds of problems: they can destroy the {\em tempered} nature of the Wightman functions
and/or they can introduce tachyonic states in the spectrum, so the modified postulate of
local commutativity is not preserved. This can be a signal that the NCFT must be analyzed
from another set of principles, whose mathematical basis may shed new light on the existence
(or not) of many of the desirable properties of fields, in particular the CPT operator and
the connection between Spin-Statistics.

Clearly, in order to formulate a NCFT, a different type of generalized function space
has to be used. Firstly, we must to keep in mind that NCFT are {\em nonlocal} due to the
commutation relations $[\widehat{x}_\mu,\widehat{x}_\nu]=i \theta_{\mu\nu}$. Secondly,
this nonlocality can have implications on highly physical properties. For example, in the
formulation of general properties of a field theory the localization plays a fundamental
role in the concrete realization of the locality of field operators in coordinate space
and the spectral condition in energy-momentum space, which are achieved through the
{\em localization of test functions} -- the fields are considered tempered functionals on
the test function space ${\mathscr S}({\Bbb R}^n)$, the Schwartz space of rapidly decreasing
functions. However, the nonlocal character of the interactions in NCFT seems to suggest that
there are good reasons to expect that fields are not tempered. Therefore, from a mathematical
point of view, we must to have serious attention with the decision about the choice of the
test function space to be used. This means that the extension of Wightman axioms to the
context of NCFT has deeper roots in the mathematical structure and must first starts with
replacement of the standard space ${\mathscr S}({\Bbb R}^n)$ by another suitable space.
As a matter of fact, since Wightman and G\r{a}rding formulated the
quantum field theory in an axiomatic way, by regarding fields as operator-valued tempered
distributions, many authors have attempted to generalize the theory to take in more fields
represented by more singular generalized functions by restricting the class of test functions.
We would like to mention the work by Jaffe~\cite{Jaffe}, on the strictly localizable field
theory, and the work by Nagamachi-Mugibayashi~\cite{NaMu} and Br\"uning-Nagamachi~\cite{BruNa}
on the field theory in terms of Fourier hyperfunctions. Wightman himself suggested that
physically relevant interacting fields do not seem to be tempered~\cite{Wight}.

The class of distributions on which a NCFT should be built has been explored some time by
L\"ucke~\cite{Luc1}-\cite{Luc3} and Soloviev~\cite{Solo1}-\cite{Solo4}. These authors have
shown that one adequate solution to treat field theories with nonlocal interactions,
it is to take the fields to be averaged with test functions belonging to the space
${\cal S}^0({\Bbb R}^n)$ consisting of the restrictions to ${\Bbb R}^n$ of entire
analytic functions on ${\Bbb C}^n$, whose Fourier transform is just the Schwartz space
${\mathscr D}({\Bbb R}^n)$ of $C^\infty$ functions of compact support. The space
${\cal S}^0({\Bbb R}^n)$ is the smallest space among the Gelfand-Shilov
spaces~\cite{GeSh}, ${\cal S}^\beta({\Bbb R}^n)$, where $0\leq \beta < 1$, which naturally
allows us to treat a theory of {\em nonlocalizable fields}. Elements in the dual space
${\cal S}^{\prime 0}$ of the space of entire functions are called analytic functionals.
Because the elements in ${\cal S}^0$ are entire functions, the locality axiom cannot be
formulated in the usual way, {\em i.e.}, there is no sensible notion of support for
distributions in ${\cal S}^{\prime 0}$. For this reason, in principle, expressing physical
requirements such as the causality in clear way become problematic. Consequently, structural
results of quantum field theories (QFT) usually seen to be firmly dependent on the locality,
and that are supposed to be universally valid, do not {\em a priori} hold to the NCFT case.
These structural results of QFT we have in mind here are ({\em i}) the existence of the
fundamental symmetry CPT, and ({\em ii}) the Spin-Statistics connection. 

The main purpose of this work is to show that, from a distributional-theoretical framework,
the analysis of \'Alvarez-Gaum\'e and V\'asquez-Mozo~\cite{AGVM} on the extension
of the Wightman axioms to the context of NCFT can be reformulated by adoption immediate
of some ideas introduced by Soloviev~\cite{Solo1}-\cite{Solo4},
which are applicable to non-commutative quantum field models. Then, we examine how
essential are the locality of interactions and the microcausality axiom in order to
reach a conclusion about the validity of the CPT and Spin-Statistics theorems in NCFT,
in the case of only spatial non-commutativity -- this choice preserves the
unitarity~\cite{Gomis} and the causality~\cite{SST}.  We show that both theorems hold within
this environment if the postulate of microcausality suggested by \'Alvarez-Gaum\'e and
V\'asquez-Mozo~\cite{AGVM} is replaced with a condition implying the macrocausality as
suggested by Soloviev~\cite{Solo1}-\cite{Solo4}. This makes our proofs stronger than ones
given by \'Alvarez-Gaum\'e and V\'asquez-Mozo~\cite{AGVM}, in the sense that our hypotheses
are weaker. 

The article is organized as follows. Section 2 contains a brief sketch of some results
by Soloviev about modern functional analysis that make it possible to extend the basic results
of axiomatic approach~\cite{SW,BLOT} to nonlocal interactions. In Section 3, the necessary
modifications of the Wightman axioms to include the case of NCFT are explained. In Section 4,
we outline the arguments that guarantee the validity of the CPT theorem and the Spin-Statistics
relation for NCFT. Section 5 contains concluding remarks. We conclude with the Appendix \ref{Ap1}
which contains the sketch of proof of the Proposition \ref{auxpro}.

\section{Analytic Functionals and Angular Localizability}
\label{Sec2}
\hspace*{\parindent}
In this section, we are going to introduce the basic facts which allow handling the analytic
functionals of class ${\cal S}^{\prime 0}$, in most cases, as easily as tempered distributions.
As mentioned above, the elements of ${\cal S}^{\prime 0}$ are analytic functionals in the
coordinate representation for which the notion of support is inapplicable. Nevertheless,
Soloviev has shown that the functionals of this class retain a kind of {\em angular localizability},
which ensures the existence of minimal carrier cones of the distributions in ${\cal S}^{\prime 0}$.
This replaces the notion of support for nonlocalizable distributions and leads to a natural
generalization of the local commutativity in NCFT with singular short distance behavior,
since the latter theories are nonlocal. For sake of completeness, we recall this property
here {\em mutatis mutandis} -- the reader is referred to~\cite{Solo1}-\cite{Solo4} and
references therein for details.

We start recalling that the space of test functions composed by entire analytic functions is
such that the following estimate holds:
\[
|f(z)| \le C_N\left(1+|x|\right)^{-N}e^{b|y|}\,\,,\quad (z=x+iy)\,\,,\quad
N=0,1,\dots\,\,,
\]
where $b$ and $C_N$ are positive constants depending on $f$. The space of functions
satis\-fying the estimate above, with fixed $b$, is denoted as ${\cal S}^{0,b}$, while in
nonlocal field theory the union $\cup_{b>0}{\cal S}^{0,b}$ is denoted as ${\cal S}^0$.
Together with space ${\cal S}^0(\oR^n)$, we introduce a space associated with closed cones
$K \subset \oR^n$. One recalls that $K$ is called a cone if $x \in K$ implies
$\lambda x \in K$ for all $\lambda > 0$.
 Let $U$ be an open cone in $\oR^n$. For each $U$, one assigns a space ${\cal S}^0(U)$
consisting of those entire analytic functions on $\oC^n$, that satisfy the inequalities
\begin{align}
|f(z)| \le C_N\left(1+|x|\right)^{-N}e^{b|y|+ bd(x,U)}\,\,,\quad
N=0,1,\dots\,\,,
\label{eq.1}
\end{align}
with $d(x,U)$ being the distance from the point $x$ to the cone $U$.\footnote{The norm in
$\oR^n$ is assumed to be Euclidean.} This space can naturally be given a topology by regarding
it as the inductive limit of the family of countably normed spaces ${\cal S}^{0,b}(U)$ whose
norms are defined in accordance with the inequalities (\ref{eq.1}), {\em i.e.},
\[
\|f\|_{U,b,N} =
\sup_z |f(z)|\left(1+|x|\right)^N e^{-b|y|-bd(x,U)}.
\]
For each closed cone $K\subset \oR^n$, one also defines a space ${\cal S}^0(K)$
by taking another inductive limit through those open cones $U$ that contain
the set $K\setminus\{0\}$ and shrink to it. Clearly, ${\cal S}^0(\oR^n)={\cal S}^0$. As usual,
we use a prime to denote the continuous dual of a space under consideration. A closed cone
$K\subset\oR^n$ is said to be a {\it carrier} of a  functional $v\in {\cal S}^{\prime 0}$ if
$v$ has a continuous extension to the space ${\cal S}^0(K)$, {\em i.e.}, belongs to
${\cal S}^{\prime 0}(K)$. As is seen from estimate (\ref{eq.1}), this property may be thought
of as a fast decrease -- no worse than an exponential decrease of order 1 and maximum type --
of $v$ in the complement of $K$. It should also be emphasized that  if  $v$ is a
tempered distribution with support in $K$, then the restriction $v|_{{\cal S}^0}$ is carried
by $K$.

We now list some results proved by Soloviev, which formalize the property of angular
localizability:
\newcounter{numero}
\setcounter{numero}{0}
\def\Theo{\addtocounter{numero}{1}\item[{$\bf R.\thenumero$}]}

\begin{enumerate}

\Theo {\em The spaces ${\cal S}^0(U)$ are Hausdorff and complete.
A set $B\subset {\cal S}^0(U)$ is bounded if, and only if, it is contained
in some space ${\cal S}^{0,b}(U)$ and is bounded in each of its norms}.

\Theo {\em The space ${\cal S}^0$ is dense in every ${\cal S}^0(U)$ and in every
${\cal S}^0(K)$}.

\Theo {\em If a functional $v\in {\cal S}^{\prime 0}$ is carried by each of closed
cones $K_1$ and $K_2$, then it is carried by their intersection}.

\Theo {\em If $v\in {\cal S}^{\prime 0}(K_1 \cup K_2)$, then $v=v_1+v_2$, where
$v_j\in {\cal S}^{\prime 0}(U_j)$ and $U_j$ are any open cones such that
$U_j\supset K_j\setminus\{0\}$, $j=1,2$}.

 \end{enumerate} 

\section{Alternative Axiomatic Description for NCFT}
\hspace*{\parindent}
The axiomatic approach proposed by G\r{a}rding and Wightman for QFT consists in studying
the consequences of a set of a few fundamental postulates on the theory such as the
relativistic invariance, the locality, the existence and stability of the vacuum, in order
to verify whether are logically consistent the basic principles suggested by the two pillars
of the modern physics: the Relativity Theory and the Quantum Mechanics. The Wightman axioms
can be summarized as follows: ({\em i}) States are described by vectors of a Hilbert space with
positive definite metric. ({\em ii}) There is a {\em vacuum state} $|\Omega_o \rangle$
with the properties of being the state of lowest energy and invariant by all the unitary
operators ${\mathscr U}(\Lambda,a)$ of the Poincar\'e group, where $a$ is a space-time
translation and $\Lambda$ is a Lorentz transformation. ({\em iii}) The local
fields are tempered distribution valued field operators. ({\em iv}) The spectrum of the
energy-momentum operator is contained in the closed forward light cone. This condition is
equivalent to the condition that the operators $p_0$ and $p^2$ are both positive.
({\em v}) Cyclicity of the vacuum. This means that one can construct a dense set of
states in Hilbert space by application of products of field operators on this state.
This condition ensures that the Hilbert space is not too large.

It turns out that these properties can be fully reexpressed in terms of an equivalent set
of properties of the vacuum expectation values of their ordinary field products, called
{\em Wightman functions} (or correlation functions of the theory):
\begin{eqnarray}
{\mathscr W}_n(x_{1},\ldots,x_{n})\overset{\text{def}}
{=}\langle\Omega_o|\Phi(x_1)\cdots \Phi(x_n)|\Omega_o\rangle\,\,.
\label{vev}
\end{eqnarray}
For most purposes, the basis
of the formulation of a QFT starts from a given set of Wightman functions which are
assumed to satisfy the following properties:
\setcounter{numero}{0}
\def\Pro{\addtocounter{numero}{1}\item[{$\tt P.\thenumero$}]}
\begin{enumerate}

\Pro Wightman functions are tempered distributions.

\Pro Wightman functions are invariant under the inhomogeneous Lorentz group.

\Pro Spectral condition: the Fourier transforms of the Wightman functions have support
in the region
\begin{gather*}
\Bigm\{(p_1,\ldots,p_n)\in {\Bbb R}^{4n}\,\,\bigm|\,\,\sum_{j=1}^{n}p_j=0,\,\,
\sum_{j=1}^{k}p_j \in {\overline V}_+,\,\,k=1,\dots,n-1 \Bigm\}\,\,,
\tag{SC}
\end{gather*}
where ${\overline V}_+=\{(p^0,{\bf p}) \in \oR^4 \mid p^2 \geq 0, p^0 \geq 0\}$ is the
closed forward light cone.

\Pro Local commutativity
\[
{\mathscr W}_n(x_{1},\ldots,x_i,x_{i+1},\ldots,x_{n})=
{\mathscr W}_n(x_{1},\ldots,x_{i+1},x_i,\ldots,x_{n})\,\,,\quad{\mbox{if}}\quad
(x_i-x_{i+1})^2<0\,\,.
\]
 \Pro Condition of positive definiteness.

\end{enumerate} 
It can be shown on general grounds that a quantum field theory which satisfies all
these conditions respect the CPT and Spin-Statistics theorems~\cite{SW,BLOT}.

As indicated in the Introduction, in NCFT the contact with the axiomatic formalism is made
by modifying some of the Wightman axioms. In effect, the most of the properties can be taken
over in parallel with Wightman approach for tempered fields, except by following
modifications:

\,\,\,{\em Modification 1}: We replace the test function space ${\mathscr S}(\oR^n)$
by ${\cal S}^0(\oR^n)$.

\,\,\,{\em Modification 2}: We replace the axiom of local commutativity, which cannot
be formulated in terms of the analytic test functions, by asymptotic commutativity in the
sense of Soloviev.

\,\,\,The main purpose of the present paper is to show that NCFT subject to these modifications
respect the CPT and Spin-Statistics theorems, within the same framework along the lines
sketched in Refs.\cite{Solo1}--\cite{Solo4}.

\subsection{Fields are Operator-Valued Distributions}
\hspace*{\parindent}
As already mentioned, for NCFT we will assume that all fields are
operator-valued generalized functions\footnote{Throughout this paper, the term generalized
functions is used synonymously with distributions.} living in an appropriate space of
functions $f(x) \in {\cal S}^0({\Bbb R}^{4n})$, the space of entire analytic test functions.
In particular, we will consider only one scalar field $\Phi(x)$. We shall denote by
$D_0$ the minimal common invariant domain, which is assumed to be dense, of the field
operators in the Hilbert space ${\mathscr H}$ of states, {\em i.e.}, the vector subspace
of ${\mathscr H}$ that is spanned by the vacuum state $| \Omega_o \rangle$ and by
various vectors of the form $ \Phi(f_1)\dots \Phi(f_n)| \Omega_o \rangle$ --
with $(n=1,2,\dots)$ -- where $f_i(x) \in {\cal S}^0(\oR^4)$. It should be noted that, the
space ${\cal S}^0$, being Fourier-isomorphic to ${\mathscr D}$, is {\it nuclear}. The
property of nuclearity enables us to define in addition the expressions
\begin{equation}
\Phi^n(f)=\int dx_1 \cdots dx_n\,\,\Phi(x_1)\cdots \Phi(x_n)\,
f(x_1,\ldots,x_n)\,| \Omega_o \rangle
\qquad(n=1,2,\ldots)\,\,,
\label{eq.5}
\end{equation}
and to verify that every operator $\Phi(f)$ can be extended to the subspace $D_1\supset D_0$
spanned by vectors (\ref{eq.5}).

\subsubsection{Vacuum Expectation Values of Fields}
\hspace*{\parindent}
The Wightman generalized functions ${\mathscr  W}_n \in {\cal S}^{\prime 0}(\oR^{4n})$,
{\em i.e.}, the vacuum expectation values of fields (\ref{vev}) define analytic distributions
on space ${\cal S}^0({\Bbb R}^{4n})$. In~\cite{Cha1}, Chaichian {\it et al.} propose new
Wightman functions as vacuum expectation values of field operators which involve the
$\star$-products
\begin{eqnarray}
{\mathscr W}_{\star}(x_{1},\ldots,x_{n})\overset{\text{def}}
{=}\langle\Omega_o|\Phi(x_1)\star\Phi(x_2)\star\dots \star\Phi(x_n)|\Omega_o\rangle\,\,,
\label{NFW}
\end{eqnarray}
where
\begin{equation*}
\Phi(x_1)\star\Phi(x_2)=e^{\frac{i}{2}\theta^{\mu\nu}
\frac{\partial}{\partial x_1^\mu}
\frac{\partial}{\partial x_2^\nu}}\Phi(x_1)\Phi(x_2)\,\,,
\end{equation*}
represents a generalization of the $\star$-product for non-coinciding points, with the
relation between the ordinary Wightman functions ${\mathscr W}_n(x_{1},\ldots,x_{n})$ and
the new functions ${\mathscr W}_{\star}(x_{1},\ldots,x_{n})$ being defined by
\begin{equation*}
{\mathscr W}_{\star}(x_{1},\ldots,x_{n})=e^{\frac{i}{2}\theta^{\mu\nu}
\sum\limits_{i<j}\frac{\partial}{\partial x_i^\mu}\frac{\partial}{\partial x_j^\nu}}
{\mathscr W}_n(x_{1},\ldots,x_{n})\,\,,
\end{equation*}
which is a consequence of
\begin{equation*}
\Phi(x_1)\star \cdots \star \Phi(x_n)=e^{\frac{i}{2}\theta^{\mu\nu}
\sum\limits_{i<j}\frac{\partial}{\partial x_i^\mu}\frac{\partial}{\partial x_j^\nu}}
\Phi(x_1)\cdots\Phi(x_n)\,\,.
\end{equation*}

The new formulation of the Wightman functions has the advantage of including explicitly
non-commutativity effects. In~\cite{AGVM} the deformation parameter $\theta$
appears only to indicate that the Lorentz invariance is ``broken'' to a lower symmetry.
However, in~\cite{Cha1} the fields are still assumed to be {\em tempered} fields. In our
opinion, same within the framework proposed in~\cite{Cha1} the fields must be considered
as generalized functions on space ${\cal S}^0({\Bbb R}^{4n})$.

In what follows, by Wightman distributions we shall understand the vacuum expectation
values of fields given by (\ref{vev}). It is very clear that the proofs of the CPT and
Spin-Statistics theorems in Section 5 hold without modifications for the distributions
(\ref{NFW}), since the $\star$-products introduced in the new Wightman functions no affect
the analytic continuation of these functions to the complex plane with respect the
``electrical'' coordinates $x_e=(x^0,x^1)$.\footnote{The coordinates
${\vec{x}}_{m}=({x}^{2},{x}^{3})$ are called ``magnetic'' coordinates in~\cite{AGVM}.}
 
\subsection{Asymptotic Commutativity}
\hspace*{\parindent}
In extending the usual Wightman framework of the axiomatic quantum field theory
our greatest concern is how to formulate the locality axiom for NCFT. The strict
localizability of fields is connected, mathematically, with the fact that the
test function space contains $C^\infty$ functions of compact support. This
requirement is satisfied by Wightman fields, because these fields were
originally constructed on the basis of the space ${\mathscr S}$. But, this is not the
case for fields constructed on the basis of the space ${\cal S}^0$.

In order to adapt the postulate of microcausality for NCFT, \'Alvarez-Gaum\'e and
V\'azquez-Mozo~\cite{AGVM} (see also Ref.\cite{AGJLB}) have relaxed the condition that
field (anti)commuta\-tors vanish outside the light cone, by replacing the light cone by
the light {\it wedge} $V_{lw}=\{(x_e,\vec{x}_m) \in \oR^{4} \mid x_e^2=0\}$
so that
\begin{eqnarray}
\bigl[\Phi(x_e,\vec{x}_m),\Phi(x_e^\prime,\vec{x}_m^\prime)\bigr]_{\pm}=0\,\,,
\quad \mbox{if} \quad (x_e-x_e^\prime)^2=(x^0-x^{\prime 0})^2-(x^1-x^{\prime 1})^2<0\,\,.
\label{LC}
\end{eqnarray}

Here, following Soloviev~\cite{Solo1}-\cite{Solo4}, we relax the condition (\ref{LC}) and
introduce, as a substitute for the axiom of locality, the following axiom:

\begin{definition}[Axiom of asymptotic (anti)commutativity]$\!\!$\footnote{It should be
mentioned that the term ``asymptotic commutativity'' was introduced within the context
which we have followed in Ref.~\cite{FaiSolo} and with some other meaning in Ref.~\cite{Luc1}.} 
The field components $\Phi(x_e,\vec{x}_m)$ and $\Phi(x_e^\prime,\vec{x}_m^\prime)$ are said
to (anti)commute asymptotically for sufficiently large space-like separation of their
arguments, if the functional
\begin{align}
f=\bigl\langle\Theta,\,\bigl[\Phi(x_e,\vec{x}_m),
\Phi(x_e^\prime,\vec{x}_m^\prime)\bigr]_\pm \Psi\bigr\rangle\,\,,
\label{AofAC}
\end{align}
is carried by the closed light wedge $\overline{V}_{e}^{\,(2)} \times \oR^4=
\{(x_e,\vec{x}_m);(x_e^\prime,\vec{x}_m^\prime) \in \oR^{4 \cdot 2} \mid (x_e-x_e^\prime)^2
\geq 0\}$ for any vectors $\Theta,\Psi\in D_0$.
\label{def3.1}
\end{definition}

A few comments about our requirement that the functional (\ref{AofAC}) is carried by
the closed wedge $\overline{V}_{e}^{\,(2)} \times \oR^4$ are now in order. In~\cite{AGVM}
this condition has been supported by the $SO(1,1)\times SO(2)$ symmetry, which is the
feature arising when one has only spatial non-commutativity~\cite{Comm}. This fact leads
the authors of~\cite{AGVM} to argue that the notion of a light cone is generally modified
to that of a light wedge. More recently, Chu {\it et al.}~\cite{Chu} have shown that the
reduction from light cone to light wedge is a generic effect for non-commutative geometry
and it is independent of the type of Lorentz symmetry breaking interaction. 

A key point of the our argument is the equality ${\cal S}^0(\oR^{4}\times
\oR^{4})={\cal S}^0(\oR^{4}) \widehat{\otimes}_i {\cal S}^0(\oR^{4})$, where the index $i$
indicates that the tensor product is endowed with the inductive topology and the hat means
the corresponding completion. By definition of the inductive topology, the dual space of
${\cal S}^0(\oR^{4}) \widehat{\otimes}_i {\cal S}^0(\oR^{4})$ is isomorphic to the
space of separately continuous functionals on ${\cal S}^0(\oR^{4}) \times 
{\cal S}^0(\oR^{4})$ (see~\cite{Solo3,Solo4}). By result $\bf R.2$ in Section \ref{Sec2},
the space ${\cal S}^0(\oR^{4})$ is dense in ${\cal S}^0(U)$, where $U$ is any open cone
in $\oR^{4}$ such that $\overline{V}_{e}^{\,(2)} \setminus \{0\} \subset U$. Hence,
the functional $f$ is carried by the closed cone $\overline{V}_{e}^{\,(2)} \times \oR^4$.
Moreover, a consideration analogous to that of Lemma 3 in~\cite{Solo3} shows that if we
introduce, for $0 \leq j \leq n$ and $n=0,1,2,\ldots$, Wightman functions
${\mathscr W}_{n,j} \in {\cal S}^{\prime 0}(\oR^{4(n+2)})$ defined by
\begin{align*}
{\mathscr W}_{n,j}(x_1,\ldots,x_j,x,y,x_{j+1},\ldots,x_n)=&
{\mathscr W}_{n+2}(x_1,\ldots,x_j,x,y,x_{j+1},\ldots,x_n)\\[3mm]
&\pm{\mathscr W}_{n+2}(x_1,\ldots,x_j,y,x,x_{j+1},\ldots,x_n)\,\,,
\end{align*}
it follows from the asymptotic commutativity condition that ${\mathscr W}_{n,j}$
defined on ${\cal S}^{0}(\oR^{4(n+2)})$ has a continuous extension to the space
${\cal S}^{0}(\oR^{4j}\times (U \times \oR^4) \times \oR^{4(n-j)})$.
Then, ${\mathscr W}_{n,j}$ is carried by the closed cone
$\oR^{4j}\times (\overline{V}_{e}^{\,(2)} \times \oR^4) \times \oR^{4(n-j)}$.

\section{Proof of the CPT and Spin-Statistics Theorems}
\hspace*{\parindent}
In this section, we want to show that the arguments above allow us to deduce some structural
results of NCFT. We have in mind here the existence of the CPT symmetry~\cite{Cha,AGVM}
and the connection between spin-statistics~\cite{Cha}. The proof of these results as given
in the literature~\cite{SW,BLOT} usually seem to rely on the tempered character of the
distributions in an essential way. In the approach which we follow the main sources of
difficulties in proving these results are: the absence of test functions of compact support
and the fact that for functionals belonging to ${\cal S}^{\prime 0}$ the notion of support
breaks down. In~\cite{Luc2,Luc3}, L\"ucke overcame this problem using the analyticity
properties of vacuum expectation values in the momentum space and analyzed the relevant
envelopes of holomorphy. More recently, an alternative and elegant
solution has been given by Soloviev~\cite{Solo1,Solo2,Solo3} using the notion of the analytic
wavefront set and a type of uniqueness theorem for distributions. For simplicity, we are
going to discuss the CPT and Spin-Statistics theorems only for the scalar field $\Phi(x)$.
Ours results are corollaries of the Soloviev's Theorems generalizing the CPT invariance and
Spin-Statistics connection for nonlocal field theories, and ours proofs follow
directly from the chain of reasoning of~\cite{Solo1}-\cite{Solo4}.

\subsection{CPT Invariance}
\hspace*{\parindent}
Let $\Phi(x)$ be a Hermitian scalar field. For this field, it is well-known that in terms
of the Wightman functions, a necessary and sufficient condition for the existence of CPT
theorem is given by:
\begin{equation}
{\mathscr W}_n(x_1,\ldots,x_n)={\mathscr W}_n(-x_n,\ldots,-x_1)\,\,.
\label{cptt1}
\end{equation}
Under the usual temperedness assumption, the proof of the equality (\ref{cptt1}) as given
by Jost~\cite{Jost} starts from the weak local commutativity (WLC) condition, namely under
the condition that the vacuum expectation value of the commutator of $n$ scalar fields
vanishes outside the light cone, which in terms of Wightman functions takes the form
\begin{equation}
{\mathscr W}_n(x_1,\ldots,x_n)-{\mathscr W}_n(x_n,\ldots,x_1)=0\,\,,\quad
{\mbox{for}}\quad (x_1,\ldots,x_n) \in {\mathscr J}_n\,\,,
\label{wlc}
\end{equation}
where $\mathscr J_n$ represents the set of Jost points, which are real points lying outside
the light cone. This implies that $(x_k-x_{k+1})^2 < 0$ for $k=1,\ldots,n-1$. Jost's proof
that the WLC condition (\ref{wlc}) is equivalent to the CPT symmetry (\ref{cptt1}) one relies
on the fact that the proper complex Lorentz group contains the total space-time inversion.
Therefore, by the BHW theorem, the equality ${\mathscr W}_n(x_n,\ldots,x_1)=
{\mathscr W}_n(-x_n,\ldots,-x_1)$ holds taking into account the symmetry property
${\mathscr J}_n=-{\mathscr J}_n$ in whole extended analyticity domain.

In order to prove that equality (\ref{cptt1}) holds in NCFT, the following auxiliary
proposition (proved in Appendix \ref{Ap1}) will be fundamental.

\begin{proposition}
The functional $F(x)={\mathscr W}_n(x_1,\ldots,x_n)-{\mathscr W}_n(-x_1,\ldots,-x_n)$,
is carried by the complement of the Jost points
\begin{equation}
{\mathscr J}_n=\Bigl\{(x_1,\ldots,x_n) \in \oR^{4n}\,\,\bigm|\,\,\Bigl(\sum_{i=1}^{n-1}
\lambda_i(x_{e_i}-x_{e_{i+1}})\Bigr)^2 < 0, \lambda_i \geq 0, \sum_{i=1}^{n-1}
\lambda_i > 0\Bigr\}\,\,.
\tag{JP}
\end{equation}
\label{auxpro}
\end{proposition}

\begin{remark}
It is worthwhile to emphasize that, according to Ref.~\cite{AGVM},
for $n>2$ the Jost points are formed by $(x_{e_i}-x_{e_{i+1}})^2<0$ with
the condition that $x_i^1-x_{i+1}^1>0$.
\end{remark}

We also formulate an analogous of the WLC condition:

\begin{definition}
The non-commutative quantum field $\Phi(x)$ defined on the test function space
${\cal S}^0(\oR^{4n})$ is said to satisfy the weak asymptotic commutativity (WAC)
condition if the functional on left-hand side of Eq.(\ref{wlc}) is carried by the
set $\complement{\mathscr J}_n$ complementary to the Jost points {\rm(JP)}.
\end{definition} 

\begin{theorem}[Modified CPT Theorem]
In a non-commutative scalar field theory in which the mo\-di\-fied Wightman axioms hold,
the weak asymptotic commutativity condition is equivalent to the CPT invariance.
\label{cpttheo}
\end{theorem}

\begin{proof}
We start assuming that the CPT invariance is fulfilled.
This implies the equality
\begin{equation}
{\mathscr W}_n(x_1,\ldots,x_n)={\mathscr W}_n(-x_n,\ldots,-x_1)\,\,,
\label{wacond}
\end{equation}
Then, one subtracts the functional ${\mathscr W}_n(-x_1,\dots,-x_n)$
from the left-hand and right-hand sides of (\ref{wacond}) in order to obtain the expression:
\begin{align*}
\Bigl[{\mathscr W}_n(x_1,\ldots,x_n)-{\mathscr W}_n(-x_1,\ldots,-x_n)\Bigr]=
\Bigl[{\mathscr W}_n(-x_n,\ldots,-x_1)-{\mathscr W}_n(-x_1,\ldots,-x_n)\Bigr]\,.
\end{align*}
By Proposition \ref{auxpro}, the difference functional on the left-hand side, denoted by
$F(x)$, is carried by set $\complement{\mathscr J}_n$. Hence, from the functional
equality above, we conclude that the weak asymptotic commutativity condition is fulfilled.
The reverse is also easily proved. If the WAC is satisfied, then the difference
${\mathscr W}_n(x_1,\ldots,x_n)-{\mathscr W}_n(-x_n,\ldots,-x_1)$
is carried by the set $\complement{\mathscr J}_n \not= \oR^{4n}$. On the other hand,
by virtue of the spectral condition~\cite{SpCond}, the Fourier transform of this difference
has support in the properly convex cone (SC), defined by Property {\tt P.3}, in Section 3.
Therefore, the CPT invariance holds identically by Theorem 4 in~\cite{Solo2}, which
asserts that ${\mathscr W}_n(x_1,\ldots,x_n)-{\mathscr W}_n(-x_n,\ldots,-x_1) \equiv 0$,
since the property of this functional of having its Fourier
transform supported by the aforementioned properly convex cone requires that each carrier
of ${\mathscr W}_n(x_1,\ldots,x_n)-{\mathscr W}_n(-x_n,\ldots,-x_1)$ cannot be
different from the whole space $\oR^{4n}$. 
\end{proof}

\subsection{Spin-Statistics Connection}
\hspace*{\parindent}
Here, let us state the theorem for a scalar
field only. The general case, mainly when gauge fields are present, deserves careful study
once we have to admit an indefinite metric, which invalidates the connection of spin
with statistics due the existence of ``scalar fermions'' as the Faddeev-Popov
ghosts~\cite{Ohta}.

\begin{theorem}[Spin-Statistics Theorem]
Suppose that $\Phi$ and its Hermitian conjugate $\Phi^*$ satisfy the weak asymptotic
condition with the ``wrong'' connection of spin and statistics. Then $\Phi(x)\Omega_o=
\Phi^*(x)\Omega_o=0$.  
\end{theorem}
 \begin{proof}
Under the hypothesis of the anomalous connection between spin-statistics, the weak
asymptotic condition implies that
\begin{equation}
W_2(\xi)+W_2^{\rm tr}(-\xi) \in {\cal S}^{\prime 0}(\overline{V}_{e} \times \oR^2)\,\,,
\quad \xi=x_1-x_2\,\,,
\label{acss}
\end{equation}
with $W_2(x_1-x_2)=\langle \Omega_o,\Phi(x_1)\Phi^*(x_2) \Omega_o \rangle$
and $W_2^{\rm tr}(x_2-x_1)=\langle \Omega_o, \Phi^*(x_2)\Phi(x_1) \Omega_o \rangle$.
Using the same arguments as in the proof of the Proposition \ref{auxpro}, we conclude that
for the regularized function $W_2^{(\Lambda)}(\xi)$, the equality $W_2^{(\Lambda)}(\xi)=
W_2^{(\Lambda)}(-\xi)$ takes place for the complement of the closed light wedge
$\overline{V}_{e} \times \oR^2$, according to BHW theorem. This implies that the difference
$W_2^{(\Lambda)}(\xi)-W_2^{(\Lambda)}(-\xi)$ is carried by $\overline{V}_{e} \times \oR^2$.
Since the unregularized difference $W_2(\xi)-W_2(-\xi)$
admits a continuous extension to the space ${\cal S}^0(\overline{V}_{e} \times \oR^2)$
as $\Lambda \rightarrow \infty$, it is possible to rewrite the condition (\ref{acss}) as
$W_2(\xi)+W_2^{\rm tr}(\xi) \in {\cal S}^{\prime 0}(\overline{V}_{e} \times \oR^2)$.
On the other hand, because of the spectral condition, in momentum space, 
$\widetilde{W}_2+\widetilde{W}_2^{\rm tr}$ has support in the convex cone
$\{(p_1,p_2) \in \oR^8 \mid p_1+p_2=0, p_1 \in \overline{V}_{+}\}$.
Hence, again by Theorem 4 in~\cite{Solo2}, we get $W_2(\xi)+W_2^{\rm tr}(\xi)\equiv 0$.
Finally, after averaging the fields with a test function, we obtain $\|\Phi^*(f) \Omega_o\|^2
+ \|\Phi(\overline{f})\Omega_o\|^2=0$, which yields $\Phi(x)\Omega_o=\Phi^*(x)\Omega_o=0$.   
\end{proof}

\section{Concluding Remarks}
\hspace*{\parindent}
In the present paper, we reexamine the recent work by \'Alvarez-Gaum\'e and
V\'asquez-Mozo~\cite{AGVM} on the extension of the Wightman axioms to the context of NCFT
under an other outlook. Our results are similar to those published in~\cite{AGVM}, but
are obtained under mildly more general assumptions. We assume a weaker version of Wightman's
axioms, where ({\em a}) fields are operator-valued generalized functions living in an appropriate
space of functions $f(x) \in {\cal S}^0({\Bbb R}^{4n})$, the space of entire analytic test
functions, ({\em b}) the local (anti)commutativity is replaced by the asymptotic variant in the
sense of Soloviev. Two profound results of the ordinary QFT, the existence of the symmetry CPT
and the Spin-Statistics connection were proved to hold for the case of a theory with
space-space non-commutativity. Here, we restrict ourselves to the simplest case,
that of a single, scalar, Hermitian field $\Phi(x)$ associated with spinless particles of mass
$m>0$.

We would like to conclude mentioning a number of questions for future research based
on the ideas of this paper:

\begin{enumerate}

\item The arguments that we have used evidently may provide new insights which will allow us
to study others structural results of QFT within the axiomatic description of NCFT, such as
the existence of the Borchers class of quantum fields, a representation of the Jost-Lehmann-Dyson
type, the Haag's theorem and so on.

\item As it was pointed out in~\cite{AGVM}, for gauge theories, in particular the
non-commutative QED (NCQED), the questions associated to the Wightman axioms and their
consequences are more involved due to the UV/IR mixing. As already mentioned, the existence
of hard infrared singularities in the non-planar sector of the theory, induced by
uncancelled quadratic ultraviolet divergences, can result in two kinds of problems: they
can destroy the {\em tempered} nature of the Wightman functions and/or they can introduce
tachyonic states in the spectrum, so the modified postulate of local commutativity
{\em is not preserved}. Nevertheless, it is worthwhile to call attention to the fact that
the non-temperedness arising from hard infrared singularities in NCFT is not a newness.
Actually, in the case of the local and covariant formulation of standard gauge quantum
field theories the necessary lack of positivity allows the occurrence of hard infrared
singularities which recall those related to the UV/IR mixing in NCFT. These results appear
to reinforce the hypothesis that the infrared issue in NCFT must be dealt
with another approach which enables the simultaneous control of infrared singularities.
This problem could be attacked with the Gelfand-Shilov spaces ${\cal S}^0_\alpha({\Bbb R}^n)$,
with $\alpha > 1$ -- in order to obtain the nontriviality of ${\cal S}^0_\alpha$ --
being a multi-index which control the infrared behavior of the Wightman
functions~\cite{Solo3,Nota1}. In this case the fields under considerations are
ultradistributions, not distributions, in variables of the momentum space. This may be an
interesting step in the proof of absence of the UV/IR mixing in NCFT and of the Spin-Statistics
theorem for NCQED.

\end{enumerate}

These topics are under investigation. We intend to report our conclusions on these issues in
forthcoming papers.

\section*{Acknowledgments.}
\hspace*{\parindent}
D.H.T. Franco would like to express his heartfelt gratitude to Professor O. Piguet
for his kind invitations at the Departamento de F\'\i sica, Universidade Federal do
Esp\'\i rito Santo (UFES). C.M.M. Polito was supported by the Brazilian agency CAPES.
The authors are specially indebted to the referee for valuable comments and
suggestions, which in fact lead to the clarification of various points. They are
very grateful for his constructive criticisms.

\appendix 
\renewcommand{\theequation}{\Alph{section}.\arabic{equation}}
\renewcommand{\thesection}{\Alph{section}}

\setcounter{equation}{0} \setcounter{section}{0}

\section{Sketch of Proof of the Proposition \ref{auxpro}} \label{Ap1}
\hspace*{\parindent} 
Due to the nuclearity property for ${\cal S}^0(\oR^{4n})$~\cite{NuclTheo},
${\mathscr W}_n$ is a multilinear functional which can be uniquely identified with
a functional ${\cal W}_n(f,g)$ in ${\cal S}^{\prime 0}(\oR^{2n} \times \oR^{2n})$,
continuous se\-pa\-ra\-te\-ly over $f \in {\cal S}^{0}(\oR^{2n})$ and
$g \in {\cal S}^{0}(\oR^{2n})$, defined by
\[
\int \prod_{i=1}^n d^2x_{e_i} d^2\vec{x}_{m_i}\,\,
{\mathscr W}_n((x_{e_1},\vec{x}_{m_1}),\ldots,(x_{e_{n}},\vec{x}_{m_{n}}))
f(x_{e_1},\ldots,x_{e_{n}})g(\vec{x}_{m_1},\ldots,\vec{x}_{m_{n}})\,\,.
\]
Taking into account the invariance under translations, we pass to the difference variables
$\xi_k$, in order to obtain a generalized function
\[
{\mathscr W}_n((x_{e_1},\vec{x}_{m_1}),\ldots,(x_{e_{n}},\vec{x}_{m_{n}}))=
W_n((\xi_{e_1},\vec{\xi}_{m_1}),\ldots,(\xi_{e_{n-1}}, \vec{\xi}_{m_{n-1}}))\,\,,
\]
in ${\cal S}^{\prime 0}(\oR^{2(n-1)}\times \oR^{2(n-1)})$, which
depends on the variables $\xi_{e_i} \in \oR^{2(n-1)}$ and $\vec{\xi}_{m_i} \in \oR^{2(n-1)}$.
Following Soloviev~\cite{Solo3}, we regularize the ultraviolet behavior of $W_n$ by
multiplying its Fourier transform $\widetilde{W}_n$ with a cutoff function. Then, with
the ``magnetic'' coordinates being held fixed, we regularize the Fourier transform of
$W_n((\xi_{e_1},\vec{\xi}_{m_1}), \ldots,(\xi_{e_{n-1}},\vec{\xi}_{m_{n-1}}))$ with an
invariant cutoff function of form $\omega_\Lambda(p_e)=\omega((p_e \cdot p_e)/\Lambda^2)
\in C^\infty_0(\oR)$, where $p_e \cdot p_e=p_0^2-p_1^2$ and such that $\omega_\Lambda(p_e)=1$
for $|(p_e \cdot p_e)/\Lambda^2|\leq 1$. The regularized distribution $W_n^{(\Lambda)}$
turns into a tempered distribution, in the ``electrical'' coordinates. Thus,
$\widetilde{W}_n^{(\Lambda)}$ has an inverse Fourier-Laplace transform
${\bf W}_n^{(\Lambda)}$. The generalized function
${\bf W}_n^{(\Lambda)}((\zeta_{e_1},\vec{\xi}_{m_1}),\ldots,(\zeta_{e_{n-1}},
\vec{\xi}_{m_{n-1}}))$, with $\zeta_{e_j}= \xi_{e_j}-i\eta_{e_j}$, is an analytic function
of $2(n-1)$ complex variables on the ``subtube'' $T_{n-1}=\oR^{2(n-1)}-iV_{e_+}^{(n-1)}$,
where $V_{e_+}$ is a subcone of the forward light cone $V_+$:
\begin{equation*}
V_{e_+}=\Bigl\{(\eta_e,\vec{\eta}_m) \in V_+\,\,\bigm|\,\,
\eta_e^2>0, \eta^0>0, \vec{\eta}_m=0\Bigr\}\,\,.
\end{equation*}
Under temperedness assumption, the function ${\bf W}_n^{(\Lambda)}$ can be analytically
continued into the extended subtube $T^{\rm ext.}_{n-1}$ by the BHW theorem, with
the continued function being covariant under the Lorentz group ${\mathscr L}_+({\oC})$ --
the complexification of $O(1,1)$ is made of similar way with
that of the Lorentz group $O(1,3)$. In particular, the transformations of
${\mathscr L}_+({\oC})$ leave the magnetic coordinates invariant (see~\cite{AGVM}
for details). Moreover, ${\bf W}_n^{(\Lambda)}$ has
$W_n^{(\Lambda)}((\xi_{e_1},\vec{\xi}_{m_1}),\ldots,(\xi_{e_{n-1}},\vec{\xi}_{m_{n-1}}))=
{\mathscr W}_n^{(\Lambda)}((x_{e_1},\vec{x}_{m_1}),\ldots,(x_{e_n},\vec{x}_{m_n}))$
as boundary value as $\eta_{e_j}\rightarrow 0$. Hence, the equality
${\mathscr W}_n^{(\Lambda)}(x_1,\ldots,x_n)={\mathscr W}_n^{(\Lambda)}(-x_1,\ldots,-x_n)$
takes place in the corresponding analyticity domain, keeping in mind that the inversion
of the four space-time coordinates is the product of the transformations
$I_{\rm st} \in {\mathscr L}_+(\oC)$ (see Eq.(3.5) in~\cite{AGVM} and a $SO(2)$ rotation
of 180 degrees. Since $T^{\rm ext.}_{n-1}$ contains the Jost points,
the tempered distribution $F^\Lambda(x)={\mathscr W}_n^{(\Lambda)}(x_1,\ldots,x_n)-
{\mathscr W}_n^{(\Lambda)}(-x_1,\ldots,-x_n)$ vanishes for the Jost points
and its restriction to ${\cal S}^0$ is carried by $\complement{\mathscr J}_n$.
By construction, $\widetilde{F}^\Lambda(p)$ coincides with $\widetilde{F}(p)$ in a
neighborhood of the origin, which enlarges indefinitely as $\Lambda \rightarrow \infty$,
and given that the space ${\cal S}^0(\oR^{4n})$ is dense in
${\cal S}^0(\complement{\mathscr J}_n)$, the unregularized generalized function
$F(x) \in {\cal S}^{\prime 0}(\oR^{4n})$ admits a continuous extension to the space
${\cal S}^0(\complement{\mathscr J}_n)$. This completes the outline of the proof.



\begin{thebibliography}{99}
\bibitem{SpCond} Since there is no non-commutativity in momentum
space, by assumption all Wightman function ${\mathscr W}_n$
satisfy the usual spectral condition so that for every $n$ the support of the
Fourier transform of ${\mathscr W}_n$ is contained in the set (SC) -- defined by
Property {\tt P.3}, in Section 3. Moreover, for a NCFT with $\theta_{\mu\nu}$ a
constant matrix, the spectral condition (SC) is strongly supported by twisted Poincar\'e
symmetry~\cite{Chai}. There has been shown that the commutation relations
$[\widehat{x}_\mu,\widehat{x}_\nu]=i\theta_{\mu\nu}$ can be interpreted in a Lorentz
invariant way and particles can be classified as in the usual commutative QFT case. As
a result of this, the spectral condition (SC) is given in terms of the invariants of the
theory.
\bibitem{Nota1} Particularly, this idea brings us back to a paper of
Moschella-Strocchi~\cite{MosStr} in which it was suggested that, because of the
non-temperedness, local and covariant formulations of gauge quantum field theories
may require the use of test functions space of the type ${\cal S}^\beta_\alpha({\Bbb R}^n)$.
\bibitem{NuclTheo} Let us recall the Lemma 2 of Ref.~\cite{Luc3} in a form using
the notion of carrier. Let $(f_1,\ldots,f_n)\rightarrow F_o(f_1,\ldots,f_n)$ be
a multilinear functional on ${\cal S}^0(\oR) \times \cdots \times {\cal S}^0(\oR)$
that is continuous in each variable separately. Then there is a unique generalized
function $F \in {\cal S}^{\prime 0}(\oR^n)$ fulfilling
\[
F_o(f_1,\ldots,f_n)=\int dx_1 \cdots dx_n\,\,F(x_1,\ldots,x_n)f(x_1)\cdots f(x_n)\,\,, 
\]
for $f(x_i) \in {\cal S}^0(\oR)$, with $i=1,\ldots,n$. Moreover, let $n^\prime < n$
and let $M=\overline{M}\subset \oR^{n^\prime}$. If, for arbitrary
$g(x_i) \in {\cal S}^0(\oR)$, with $i=n^\prime+1,\ldots,n$, $M$ is a carrier of
${\cal S}^{\prime 0}(\oR^{n^\prime})$ element
\[
f \longrightarrow \int dx_1 \cdots dx_n\,\,F(x_1,\ldots,x_n)f(x_1,\ldots,x_{n^\prime})
g(x_{n^\prime+1})\cdots g(x_n)\,\,, 
\]
then $M \times \oR^{n-n^\prime}$ is a carrier of $F \in {\cal S}^{\prime 0}(\oR^n)$.
\end{thebibliography}
\end{document}